\documentclass[a4paper,fleqn,usenatbib]{mnras}

\usepackage{newtxtext,newtxmath}

\usepackage[T1]{fontenc}
\usepackage{ae,aecompl}
\usepackage{graphicx}	
\usepackage{amsmath}	
\usepackage{amssymb}	
\usepackage{hyperref}
\usepackage{microtype}
\usepackage{verbatim}
\usepackage{graphicx}
\usepackage{url}
\usepackage{soul}

\title[Soft X-ray residuals in Holmberg IX X-1]{\centering{Observational limits on the X-ray emission from the bubble nebula surrounding Ho IX X-1}}

\author[R. Sathyaprakash et al.]{
Rajath Sathyaprakash,$^{1}$\thanks{E-mail: Rajath.Sathyaprakash@durham.ac.uk}
Timothy P. Roberts$^{1}$
Magdalena M. Siwek $^{2}$
\\
$^{1}$Centre for Extragalactic Astronomy, Department of Physics, Durham University, South Road, Durham DH1 3LE, UK\\
$^{2}$Harvard-Smithsonian Center for Astrophysics, 60 Garden St., Cambridge, MA 02138, USA
}

\date{Accepted XXX. Received YYY; in original form ZZZ}

\pubyear{2018}

\begin{document}
\label{firstpage}
\pagerange{\pageref{firstpage}--\pageref{lastpage}}
\maketitle

\begin{abstract}
\begin{centering}
Optical and radio observations of shock-ionised bubble nebulae surrounding ultraluminous X-ray sources (ULXs) suggest that they are powered by jets or super-critical outflows presumably launched from the ULX accretion disc. Recent simulations of these systems have shown that the shocked wind can emit thermal X-rays with estimated luminosities $\lesssim 10^{36}$ erg s$^{-1}$. In this work, we investigated whether it is possible to detect and spatially resolve the X-ray emission from these systems using archival \textit{Chandra} observations of the ULX Holmberg IX X-1. This source is an ideal target to study for two reasons: it is surrounded by an optical bubble nebula with a large spatial extent ($\sim$ 400 pc) that can easily be resolved with \textit{Chandra}. Further, it has a hard X-ray continuum that is easily distinguishable from the expected soft thermal emission from the nebula. However, a spectral and photometric analysis on stacked \textit{Chandra} observations of the source reveals that there is no strong evidence for an X-ray bubble associated with it, to a limiting luminosity of $\sim 2 \times 10^{36}$ erg s$^{-1}$. The detection of such X-ray nebulae may be possible with future X-ray missions such as \textit{Athena}, which would provide useful constraints on the kinematics of the outflow. Finally, our observations also emphasise that the nebular emission does not contribute significantly to the residuals in the X-ray spectrum of the source, which are more likely to be linked to processes localised to the ULX. 
\end{centering}
\end{abstract}


\section{Introduction}

Ultraluminous X-ray sources (ULXs) are extragalactic, off-nuclear point sources with apparent (0.3-10) keV luminosities in excess of the Eddington limit ($L_{\text{Edd}}$) for a 10 M$_{\odot}$ black hole \citep*{1}. Whilst it was initially suggested that ULXs are powered by sub-Eddington accretion onto intermediate mass black holes (IMBHs), with masses in the range (10$^{2}$ - 10$^{4}$) M$_{\odot}$, recent studies with \textit{XMM-Newton} and \textit{Chandra} (complemented by multi-wavelength observations) have argued against this hypothesis in most cases. Instead, ULXs are powered by less massive compact objects, either stellar mass black holes (StMBHs) or neutron stars (NSs), accreting close to or above $L_{\text{Edd}}$ (\citealt{2}; \citealt{3}; \citealt{4}; \citealt{5}). 

The spectral properties of ULXs yield insights into the geometry of the accretion flow surrounding the compact object. The \textit{XMM-Newton} spectra of bright ULXs (with $L_{\text{X}} > 3 \times 10^{39}$ erg s$^{-1}$; \citealt{SRM}) tend to feature two distinct thermal components. In some sources, the soft component clearly dominates over its hard counterpart (or vice versa), whilst in other cases we see the relative strengths of the two components varying for a given source (\citealt{SRM}). This behaviour is well explained by a unified model of super-Eddington accretion (\citealt{7}; \citealt{8}). The key prediction of this model is that StMBHs or low magnetic field NSs accreting above $L_{\text{Edd}}$ launch strong radiatively driven winds from an advection dominated accretion flow. There is conclusive observational evidence for such outflows in at least four ULXs (see \citealt{outflow_discovery}, \citeyear{outflow} and \citealt{kosec}). The outflow scatters and re-processes radiation from the inner-most regions of the accretion flow, which emit hard X-ray photons with a large colour correction (\citealt{10}; \citealt{fcorr_theory}). These photons are Compton down-scattered to lower energies by optically thick plasma in the outflow, which gives rise to the soft thermal component of the energy spectrum (\citealt{middleton2011}). As the mass accretion rate increases, the wind is launched from progressively larger radii where the escape velocities are smaller, thus resulting in more material being expelled from the disc. This leads to an enhanced scattering of the hard emission and it is geometrically beamed towards observers who view the system face-on, providing a natural means for the apparent luminosity to far exceed $L_{\text{Edd}}$ (\citealt{12}). On the other hand, for observers at very high inclinations the outflow obscures the hard emission from the hot inner disc, so they see a dominance of the soft thermal component. Finally, moderately inclined observers may experience transitions between these two regimes. In this way, both the accretion rate as well as the inclination angle of the system can explain the observed spectrum and its evolution (\citealt{13}). 

\begin{figure*}
\includegraphics[scale=0.5]{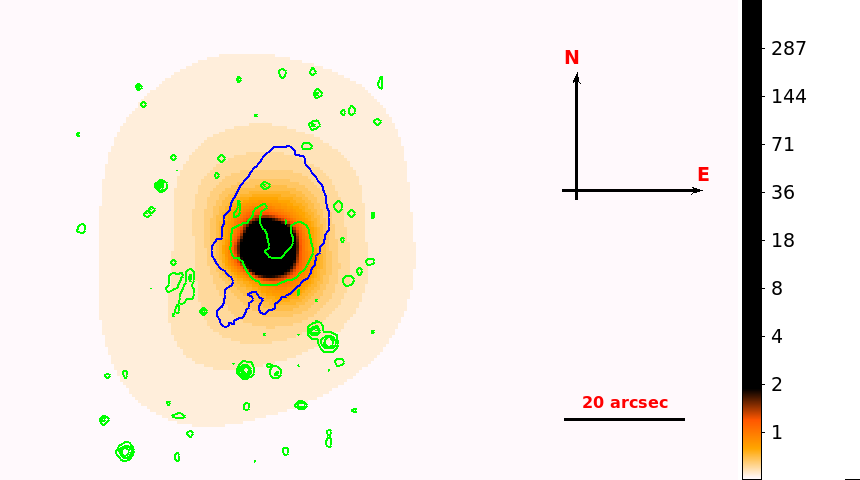}
\caption{A stacked \textit{Chandra} image of all five observations described in Table 1, with the contours of H$\alpha$ emission (from the \textit{SUBARU} FOCAS archive) overlaid. The colour-bar shows the smoothed X-ray counts per pixel in the (0.4 - 1.0) keV energy band (see text). The contour displayed in blue demarcates the edge of the expanding bubble and has a diameter of 37.6 and 24.5 arcsecs across its major and minor axis respectively, fully enclosing the brightest part of the PSF (see text). The image has been adaptively smoothed with {\tt{csmooth}} \citep*{smoothing}, at a minimum significance of 3$\sigma$ above the background.}
\label{image}
\end{figure*}

\begin{table*}
\centering
\begin{tabular}{c c c c c c}
\hline \hline
ObsID & Exposure Time (ks) & Off-axis angle (arcsec) & Frame Time (s) & Detected count rate (photons/frame time) & Pileup fraction \\[0.5ex]
\hline
	
4752& 4.96& 5.80& 1.8& 0.67 $\pm$ 0.02& 0.21 \\
4752& 5.04& 5.80& 1.8& 0.64 $\pm$ 0.02 & 0.19  \\
9540& 25.72& 130.5& 3.2& 1.37 $\pm$ 0.01& 0.39 \\
13728& 13.67& 0.25& 0.4 & 0.278 $\pm$ 0.004 & 0.10 \\
14471& 12.76& 0.25& 0.4 & 0.496 $\pm$ 0.004 & 0.18 \\[1ex]
\hline
\end{tabular}
\label{table:observations}
\caption{The five archival \textit{Chandra} observations of Ho IX X-1 utilised in this study, with a combined exposure time of 62.15 ks. The ObsID column gives the observation identifier for each source. We also quote the off-axis angle of Ho IX X-1 in each exposure, assuming the WCS coordinates for the ULX given in Gladstone et al. (2013). The pile-up fractions specified in the final column are lower limits to the true values, since the \textit{detected} (not the intrinsic) count rates of the source were used in calculating them. We calculated this with the PIMMS software provided by the \textit{Chandra} Proposal Planning Toolkit.}
\end{table*}

However, the two thermal components do not describe the data sufficiently well below 2 keV, since the \textit{XMM-Newton} spectra of several ULXs clearly show residuals at these energies (\citealt{spec_hard_inclination}). These features are resolved into a complex of narrow absorption and emission lines in four particular sources (see \citealt{outflow_discovery}, \citeyear{outflow} and \citealt{kosec}). The absorption lines are highly blue-shifted (i.e. $\frac{v}{c} \sim 0.2$) with respect to the rest-frame energies of the relevant atomic transitions. It is clear that they are produced by highly ionised atomic species in a fast-moving outflow, which strengthens the evidence for super-Eddington accretion. However, in sources with very weak detections of these features (e.g. Holmberg IX X-1; \citealt{hoixx1_outflow}), it is unclear whether the residual emission is definitively associated with the super-critical wind. Previous studies have invoked several possibilities to account for their origin. For instance, \citet*{star_formation} posited that star-formation related diffuse emission in the host galaxy of ULXs may explain the residuals in their spectra. Star-forming galaxies possess significant amounts of hot tenuous plasma thought to be heated by the combined effect of supernovae and winds from young massive stars (\citealt{star_formation3}). The soft residuals in the CCD spectra of ULXs are also well fitted by a thermal plasma emission model with an underlying bremsstrahlung continuum. However, the luminosity in this component is far too large to be solely associated with star-formation related diffuse emission local to the ULX (\citealt{residual_features}). The evidence for this was strengthened by a spatial analysis of the source NGC 5408 X-1 with \textit{Chandra} and HST observations, which clearly indicate that it is displaced from the major sites of star-formation in its host galaxy (\citealt{spatial_analysis}). Similarly, it has also been argued that the residuals are unlikely to be explained by collisional excitations of the low-density material in the nebulae surrounding some ULXs (\citealt{spec_hard_inclination}; \citealt{outflow_discovery}), although this has not yet been confirmed observationally.

ULX bubble nebulae (ULXBs) were first discovered in narrow-band optical images by \citet{ULXB_discovery}. The optical spectra of some of these sources feature strong He II $\lambda4686$ emission lines, indicating that they are photo-ionised by the X-ray continuum of the ULX. In other sources, the prevalence of strong [O I] and [S II] emission lines, with large flux ratios of the latter with respect to the Balmer H$\alpha$ line, implies that the nebula is shock-ionised (analogous to the spectra of standard supernova remnants). The evidence for this is strengthened by the the relatively large line-widths of the putative emission lines, indicating super-sonic bubble expansion velocities in the range (50-200) km s$^{-1}$ (\citealt{SNR}; \citealt*{hoII_nebula}; \citealt{hoII_nebulaOIV}). Given the association of several ULXs with young star clusters, it was initially suggested that these bubbles are powered by multiple supernova explosions of massive early-type stars in such clusters. However, further studies have argued that that this alone is insufficient to account for the mechanical energy of ULXBs ($\gtrsim 10^{52}$ ergs; \citealt{SNR}; \citealt{nebula_power}), and that their dynamics may be better explained by continual energetic input from jets or super-critical wide angle outflows launched from the ULX accretion disc \citep*{beambags}. 

The evolution of wind-inflated bubbles has been the subject of several studies (\citealt{bubble_theory1}; \citealt{bubble_theory2}). The hypersonic outflow (launched from the ULX accretion disc) shocks the surrounding ISM gas and expands adiabatically at very early times, until the mass of the swept-up material becomes comparable to the mass of the ejected outflow. This is followed by the 'snow-plough' expansion phase, when the swept-up gas starts to become radiatively efficient and collapses into a thin dense shell just behind the forward shock. Beneath this cold layer of material lies a more tenuous layer of hot shocked ejecta, which is continually heated by the mechanical energy of the outflow to temperatures in excess of $10^{7}$ K. As a result, it radiates as an optically thin plasma, producing line emission in the X-ray band with an underlying bremsstrahlung continuum. This picture has been verified in recent numerical simulations of wind-inflated ULXBs by \citet{nebula_sim}, who estimate a maximum X-ray luminosity from these sources of $\sim 10^{36}$ erg s$^{-1}$. In this work, we investigate whether it is possible to detect the X-ray emission from the shocked ejecta using archival \textit{Chandra} observations of the ULX Ho IX X-1, and verify whether this makes a significant contribution to the spectral residuals of the source.

Ho IX X-1 resides at a distance of 3.6 Mpc in an irregular dwarf galaxy (\citealt{HoIX_distance}), and consistently radiates with an X-ray luminosity above $\sim 10^{40}$ erg s$^{-1}$ (\citealt{HoIX_spectral_var}). We choose to study this source primarily because it is surrounded by a large [300 pc $\times$ 470 pc] shock-ionised nebula (\citealt{HoIX_optical_counterpart}) that is spatially resolved with \textit{Chandra}. Moreover, the source has a notably hard X-ray spectrum (\citealt{spec_var_walton}) that should easily be distinguishable from the likely soft nebular emission on larger scales. In addition, amongst those ULXs with both a strong detection of spectral residuals and a spatially resolved bubble, Ho IX X-1 has the longest exposure \textit{Chandra} data. This paper is organised as follows: we describe the spatial analysis in section 2. The spectral properties of the source are examined in section 3, particularly focusing on searching for features of thermal plasma emission from the nebula. We then confirm whether the source is actually point-like at the spatial resolution of \textit{Chandra} in section 4. Finally, we discuss the implications of our results and provide concluding statements in section 5.  

\section{Spatial analysis}

The \textit{Chandra} data were reprocessed and reduced using CIAO (v4.9)\footnote{http://cxc.harvard.edu/ciao/}. We first performed a visual check for any diffuse emission surrounding Ho IX X-1 by stacking the five observations listed in Table 1. We restricted the energy range of each observation to (0.4 - 1.0) keV. This was done in order to optimise the search for extended emission from an optically thin thermal plasma, whose luminosity is expected to peak close to a temperature of $\sim 0.3$ keV (Siwek et al. 2017). We do not consider events above 1 keV, since the power-law emission from the ULX dominates over the thermal plasma emission above this energy. 

The individual aimpoints of each observation slightly differ from one another such that they do not all have identical positions in detector coordinates. We corrected for these offsets by shifting the positions of each exposure to a common tangent point via the \texttt{reproject\_obs} task. We then checked for any external sources of contamination in the vicinity of the point source. Specifically, we removed ACIS readout streaks \citep*{readout_streak} present in three of the five exposures (ObsIDs: 13728, 14471 and 9540) by running \texttt{acisreadcorr}. The latter was provided with a spectrum of the background extracted 50 arcseconds away from the position of the ULX. The algorithm removed any events that were not consistent with this spectrum, within the region containing the readout streak events. The cleaned event file was then adaptively smoothed by \texttt{csmooth} to produce the image shown in Figure 1, binned to native ACIS pixel resolution. The overlaid contours of H$\alpha$ emission from \textit{SUBARU} FOCAS\footnote{http://smoka.nao.ac.jp/} map the position of the ULXB relative to the X-ray source. We performed astrometric corrections of point sources in the \textit{SUBARU} image by cross-matching their positions with seven sources from the USNO A2.0 catalogue\footnote{http://tdc-www.harvard.edu/catalogs/ua2.html}. This enabled us to account for any relative offset between the H$\alpha$ nebula and the ULX. We note that the astrometric error was $\sim$ 0.5 arc-seconds, which is very close to the spatial resolution of \textit{Chandra}. 

The stacked image in Figure 1 plainly does not suggest a strong evidence for any extended emission in the vicinity of the ULX. The X-ray emission appears point-like, with the regions surrounding the bubble nebula making a minor contribution to the total number of events. Prior to analysing the X-ray spectrum of the extended region, we ran the source detection algorithm \texttt{wavdetect} on the stacked image to ensure that no other point sources were detected within the extent of the bubble nebula. This would otherwise contaminate the spectrum in this region. The source detection is performed by correlating the input image with Mexican Hat functions at several scale sizes (i.e. 1,2,4,8 and 16 in pixel units). We used the default (CIAO) value for the detection significance parameter (\texttt{sigthresh} = 1 $\times$ 10$^{5}$), which ensures no more than 1 false detection over the number of pixels in the image. As such, we found that the stacked image does not contain any resolved point sources within the H$\alpha$ nebula aside from Ho IX X-1.     

\section{X-ray spectral properties}

\begin{table*}
\centering
\begin{tabular}{c c c c c c}
\hline \hline
Spectral region & XSPEC model & $N_{\text{H}}$ [10$^{22}$ cm$^{-2}$] & $\Gamma$ & $kT_{\text{p}}$ or $kT_{\text{in}}$ (keV) & $\chi^{2}$/dof. \\[0.5ex]
\hline
ULX & \tiny{tbabs $\times$ tbabs $\times$} & 0.157 $\pm$ 0.01 & 1.224 $\pm$ 0.02 & N/A & 459.7/386 \\
 & \tiny{(powerlaw)} & & & & \\
 & & & & &\\
ULX & \tiny{tbabs $\times$ tbabs $\times$} & 0.41 $\pm$ 0.1 & 1.331 $\pm$ 0.05 & 0.138 $\pm$ 0.01 & 437.2/384 \\
 & \tiny{(powerlaw + diskbb)} & & & & \\
 & & & & & \\
Extended region & \tiny{tbabs $\times$ tbabs $\times$} & 0.108 $\pm$ 0.06 & 1.29 $\pm$ 0.1 & N/A & 75.1/74 \\
 & \tiny{(powerlaw)} & & & & \\
 & & & & & \\
Extended region & \tiny{tbabs $\times$ tbabs $\times$} & 0.102$_{-0.03}^{+0.3}$ & 1.24$_{-0.1}^{+0.2}$& 0.138$^{*}$ & 75.1/73 \\
 & \tiny{(powerlaw + diskbb)} & & & & \\
& & & & & \\[1ex]
\hline
\end{tabular}
\label{table:spec_fit}
\caption{A summary of \texttt{xspec} models employed in fits to the combined ULX and extended source spectra. The quoted errors are at the 90 percent confidence level. Note that $\Gamma$ is the photon index, $N_{\text{H}}$ is the variable hydrogen column density, $kT_{\text{in}}$ is the inner disc temperature (relevant for fits with the {\tt{diskbb}} model) and $kT_{\text{p}}$ is the temperature of the {\tt{apec}} plasma. We were unable to constrain the temperature of either the {\tt{diskbb}} or the {\tt{apec}} models fit to the extended source spectrum, owing to a small number of counts at energies below 0.5 keV. So, these were fixed to the values indicated with an asterisk.}
\end{table*}
 
The X-ray emission in Figure 1 appears to be point-like, and therefore does not suggest a strong evidence for the presence of extended nebular emission. However, the radial profiles of some individual point sources indicate a hint of excess emission above the background beyond a 20 arcsecond radius (which is larger than the spatial extent of the ULXB; see Figure 5). Here, we analyse whether this is associated with the bubble nebula or is simply the emission in the PSF wings scattered off the \textit{Chandra} mirrors. In particular, we investigate whether the spectrum of the extended region (as defined in Figure 2) contains some contribution from optically thin thermal plasma emission, which would support the former scenario. The X-ray emission from the shocked wind is not expected show significant variability over timescales as short as the duration between the first and the last \textit{Chandra} observation. Therefore, we combine the spectra from all five exposures in order to enhance the signal-to-noise ratio (see below). One expects a greater (relative) contribution from the nebular emission in the wings of the PSF (where the ULX emission is fainter) than in the core, so we investigate whether this is satisfied by comparing the spectra from these two regions. 

\subsection{Spectral extraction}

We proceeded to extract the ULX spectrum inside a circular aperture of 4 pixel radius for all exposures, as recommended for a point source in the relevant \textit{Chandra} science thread\footnote{http://cxc.harvard.edu/ciao/threads/pointlike/}. The spectrum in the extended region was extracted using an annular aperture as shown in Figure 2. We set the outer radius of the annulus to 30 arcsec, ensuring that it extends well beyond the edge of the nebula, and the inner radius to the 98 percent encircled energy of the PSF. The motivation for this was to make sure that the extended source spectrum contains as little emission from the ULX as possible. We calculated the inner radius for each observation by making a curve-of-growth of the source events in a 5 arcsec circular aperture centered on the ULX. All the spectra were created by running \texttt{specextract} in CIAO, which automates the creation of the ancillary response files (ARFs) and the response matrix files (RMFs). The background spectra were extracted within circular apertures of 20 arcsecond radius on the same chip as the X-ray source (but placed sufficiently far away to avoid any contamination from it), and were subtracted from the source events. When creating the ULX spectra, we did not generate weighted RMFs (\texttt{weight=no}) and corrected for events falling outside the extraction aperture (\texttt{correctpsf=yes}), as required for point source analysis. However, when extracting the extended source spectra the RMFs were weighted (\texttt{weight=yes}), in order to account for the spatial variation in the ACIS-S detector response. For each region, we stacked the spectra of all five observations and ensured that there were a minimum of 20 counts per energy bin, which validates the use of the $\chi^{2}$ statistic in our fits.    

\begin{figure}
  \centering
  \includegraphics[width=\columnwidth]{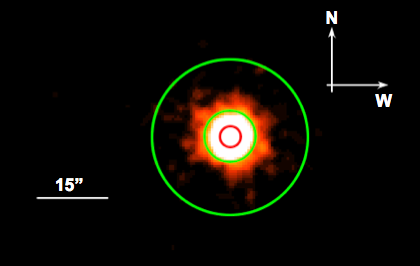}
	\caption{Stacked ACIS-S (0.3 - 6.0) keV image of Ho IX X-1 smoothed with a 2 arcsec Gaussian for display purposes. The red circle (with 4 pixel radius) marks the extraction region for the ULX spectrum, while the extended source spectra are extracted within the green annulus. The inner radius of this annulus varies for each observation, and corresponds to the radius demarcating the PSF core from the wings. The outer radius is fixed to 30 arcsec, which extends well beyond the edge of the bubble nebula (compare with Figure 1).}
  \label{spec_extract}
\end{figure} 

\subsection{A comparison of the ULX and extended emission}

\begin{figure*}
  \begin{minipage}{\textwidth}
    \centering
    \includegraphics[width=1.0\textwidth]{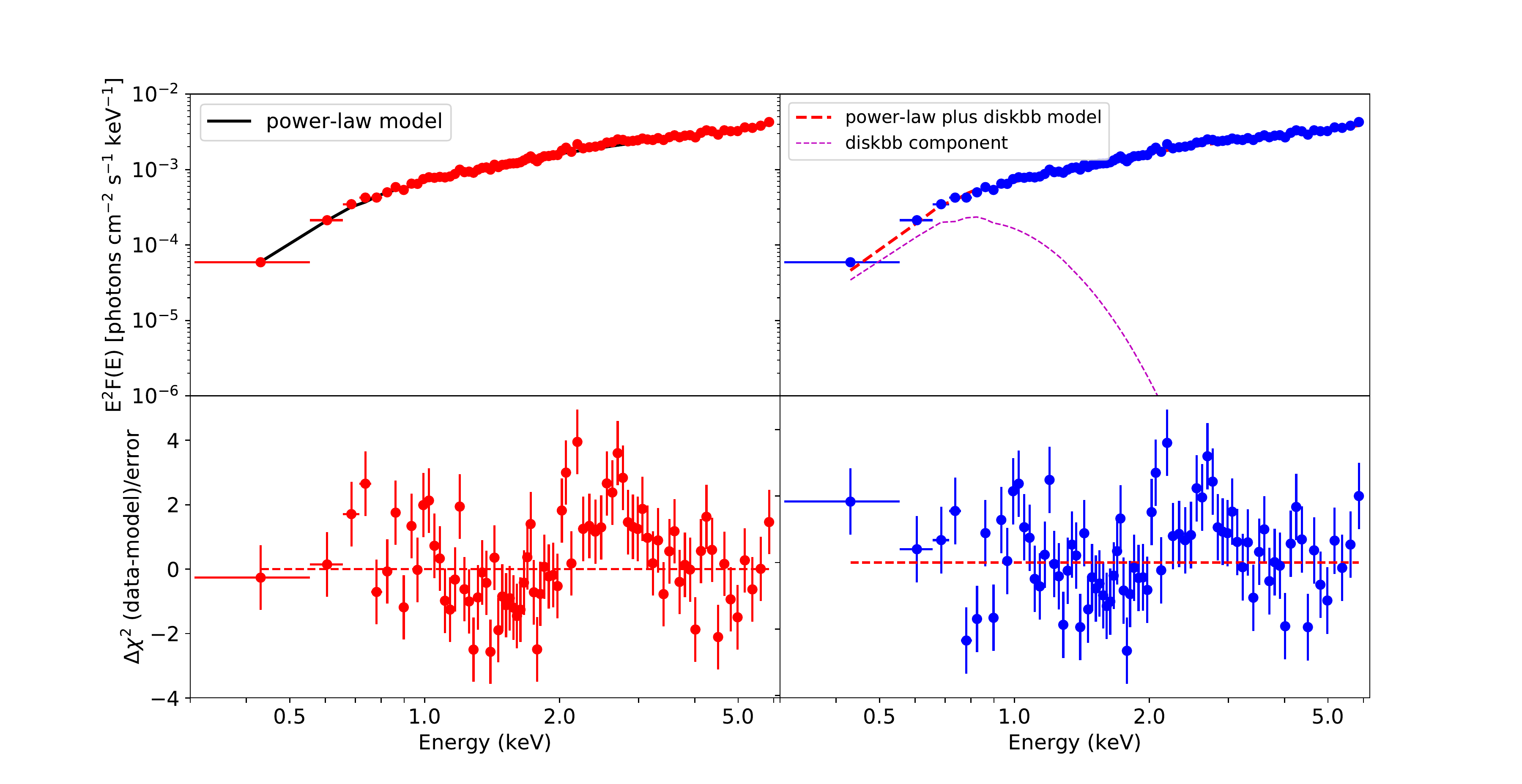}
    \caption{(Left): A stacked spectrum of the ULX, with the best-fit doubly absorbed power-law model plotted over the (0.3 - 6.0) keV energy range. A satisfactory fit to the data is obtained, with the best-fit parameters quoted in Table 2. (Right): The same spectrum now fitted with a doubly absorbed {\tt{diskbb}} plus {\tt{powerlaw}} model. This improves the goodness-of-fit significantly over just the single power-law component (see text).}
    \label{results:ULX}
  \end{minipage}\\[1em]
\end{figure*}    

\begin{figure*}
  \begin{minipage}{\textwidth}
    \centering
    \includegraphics[width=1.0\textwidth]{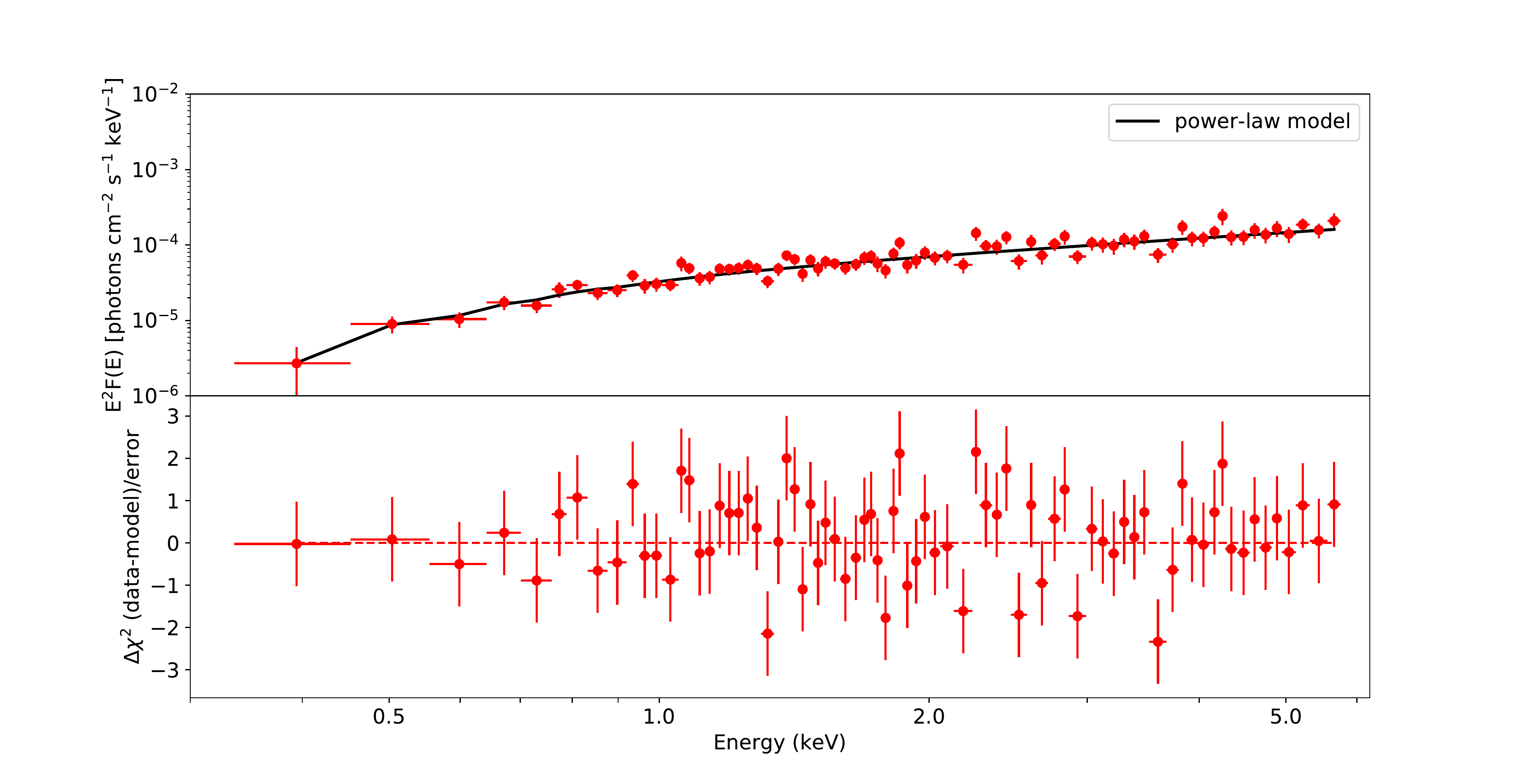}
    \caption{The stacked extended source spectrum extracted in the wings of the PSF and fitted with a doubly absorbed power-law model. The spectrum shows no evidence of any excess at energies below $1$ keV, which would be expected if the emission from the shock-ionised nebula was significant.}
    \label{results:Extended}
  \end{minipage}\\[1em]
\end{figure*}    

We began by fitting the stacked ULX spectrum with a doubly absorbed power-law model (i.e. \texttt{tbabs} $\times$ \texttt{tbabs} $\times$ \texttt{powerlaw}) in \texttt{xspec} using the abundances of \citet{abund}, and restricted the energy range of the fit to (0.3 - 6.0) keV. We included two neutral absorption components, the first of which was fixed to the Galactic value in the direction of Ho IX X-1 (5.54 $\times 10^{20}$ cm$^{−2}$; \citealt{Nhabund}), while the second was chosen to account for any absorption intrinsic to the source and/or its host galaxy. This was allowed to vary together with the remaining parameters of the model. 

A single power-law component alone did not provide a good fit to the stacked ULX spectrum, which shows some residuals at energies below 1 keV. However, the fit was substantially improved by the addition of a multi-colour disc model (i.e. $\Delta \chi^{2}$ = 22.5 for 2 additional degrees of freedom), which accounts for most of the residual emission. Previous analyses of the broadband spectrum of Ho IX X-1 have found a similarly strong evidence for the presence of a soft thermal component, which is thought to be emitted by the inner photosphere of a super-critical outflow. The peak emission temperatures found by those authors is in reasonable agreement with the value reported in Table 2 (i.e. $\sim 0.1$ keV). However, some \textit{XMM-Newton} spectra (with sufficiently high signal-to-noise) show prominent residuals even after the inclusion of a soft thermal component (\citealt*{ultraluminous}; \citealt{8}). The residuals in these spectra are well modelled by emission from collisionally ionised plasma. But, there are no strong indications of such features in the current \textit{Chandra} spectrum of Ho IX X-1. Namely, the addition of a thermal plasma emission model (on top of the pre-existing power-law and multi-colour disc components) leads to an insignificant change in the goodness-of-fit (i.e. $\Delta \chi^{2}$ = 8 for 2 additional degrees of freedom). 

We proceeded to check whether there is a stronger contribution from thermal plasma emission in the extended region by repeating the steps of analysis described above. However, the extended source spectrum appears to be featureless (Figure 4). It is very well fitted by a simple power-law model that cannot be rejected above a significance of 1$\sigma$. In this case, there is no strong requirement for an additional \texttt{diskbb} component, and the spectrum is dominated by the hard emission from the ULX (see Table 2). This is consistent with the fact that the photon indices of the ULX and extended source spectra are fully within error. Even though the radial profiles of the individual observations of Ho IX X-1 indicate some excess emission above the background in the extended region, this is unlikely to be attributed to emission from an X-ray nebula, since there are no strong features of thermal plasma emission in the extended source spectrum. Rather, it is likely to be dominated by the scattered emission of the ULX in the wings of the PSF. As an aside, we note that since four of the five \textit{Chandra} exposures suffer from pile-up (see Table 1), the accuracy of the best-fit parameters in Table 2 cannot be guaranteed. But, this does not affect the main goal of the analysis performed here. That is, we are not interested in characterising the spectral properties of the source, but rather in searching for features of diffuse emission from the bubble nebula in the PSF wings, where pile-up is negligible.       

We set an upper limit on the nebular emission by fitting the extended source spectrum with three additive model components. Namely, the (absorbed) \texttt{powerlaw} plus \texttt{diskbb} model sets the baseline for the soft emission in the extended region (which is dominated by the ULX), and we add a thermal plasma emission component \citep[\texttt{apec};][]{apec} to calculate any excess nebular emission. Owing to an insufficient number of counts at energies below 1 keV, we could not constrain the plasma temperature in our fits. However, the simulations of bubble nebulae (Siwek et al. 2017) estimate plasma temperatures from the shocked wind to be $\lesssim 0.5$ keV. So, we fixed the temperature to a range of different values between (0.2 - 0.5) keV and computed the upper limit separately for each value. As such, the largest limiting luminosity of the {\tt{apec}} component (in the 0.1 - 10.0 keV energy range) was found to be $10^{36}$ erg s$^{-1}$ (corresponding to a plasma temperature of 0.4 keV), which constitutes less than 5 percent of the total flux. As a caveat, we note that the upper limit was calculated by extrapolating the {\tt{apec}} model flux to energies below 0.3 keV, since the bandpass of \textit{Chandra} does not extend below this energy. The observed (0.3 - 10.0) keV upper limit is slightly lower (i.e. $\sim 5 \times 10^{35}$ erg s$^{-1}$). This is reasonably consistent with the estimates from Siwek et al. (2017), although we caution that they only compute this at a single energy (2 keV).       

\begin{figure*}
  \begin{minipage}{\textwidth}
    \centering
    \includegraphics[width=0.95\textwidth]{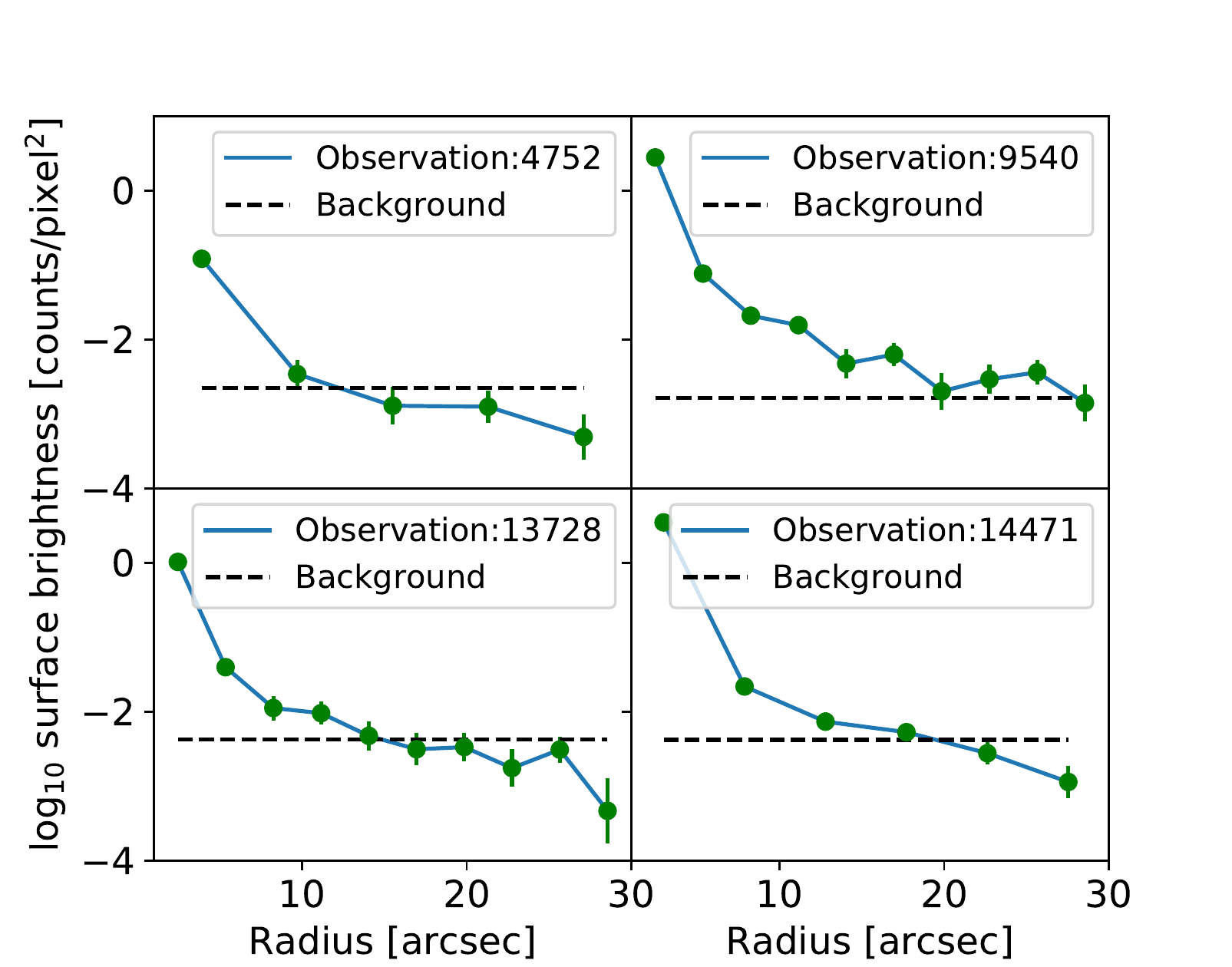}
	  \caption{Surface brightness profiles of the four longest exposure \textit{Chandra} observations of Ho IX X-1. The surface brightness stays below the background level beyond a radius of 20 arcsec in three of the four cases (see text). We have not shown the radial profile of Obs 4751, since it displays virtually the same behaviour as Obs 4752.}
    \label{radial_profiles}
  \end{minipage}\\[1em]
\end{figure*}    

\section{Photometry}

As a final test of whether Ho IX X-1 is truly point-like at the spatial resolution of \textit{Chandra}, we examined the radial profile of each observation (Figure 5). We calculated these by using a stack of concentric annuli centered on the source, out to a radius of 27 arcsecs. Figure 5 illustrates that in three of the four cases, the surface brightness does not significantly exceed the background level beyond a radius of $\sim 15$ arcsec, which is well within the extent of the ULXB. The only exception to this is Obs 9540. This observation has the largest offset from the nominal aimpoint of the instrument (see Table 1), which broadens the PSF thus giving rise to a greater number of counts in the PSF wings. However, it still does not show any excess emission in the vicinity of the bubble.  

We extended this analysis by comparing the observed PSF with a realistic (ray-tracing) simulation of the \textit{Chandra} point source via the MARX (Model of the AXAF \footnote{The \textit{Chandra} X-ray Observatory (CXO) was previously known as Advanced X-ray Astrophysics Facility (AXAF)} Response to X-rays; \citealt*{MARX}) software suite (v5.3.2). We selected the observation that was least piled-up (i.e. Obs 13728; Table 1) to compare the observed number of counts in the wings of the PSF (i.e. in the annular aperture in Figure 2) with the corresponding MARX simulation. As inputs to the latter, we provided the off-axis angle of the observation and its energy spectrum, both of which influence the encircled energy fraction of the PSF. In summary, we find that the predicted number of counts in the PSF wings (250 $\pm$ 20) is in very good agreement with the observed counts (240 $\pm$ 20). Given that Obs: 13728 is the faintest of the five exposures and the most on-axis, one expects the diffuse nebular emission to be most strongly detected in this exposure due to a comparative reduction in the number of counts from the ULX. But, this does not appear to be the case, implying a point-like nature of the source and the lack of an extended X-ray nebulosity.

\section{Discussion and conclusion}

\begin{figure*}
  \begin{minipage}{\textwidth}
    \centering
    \includegraphics[width=1.0\textwidth]{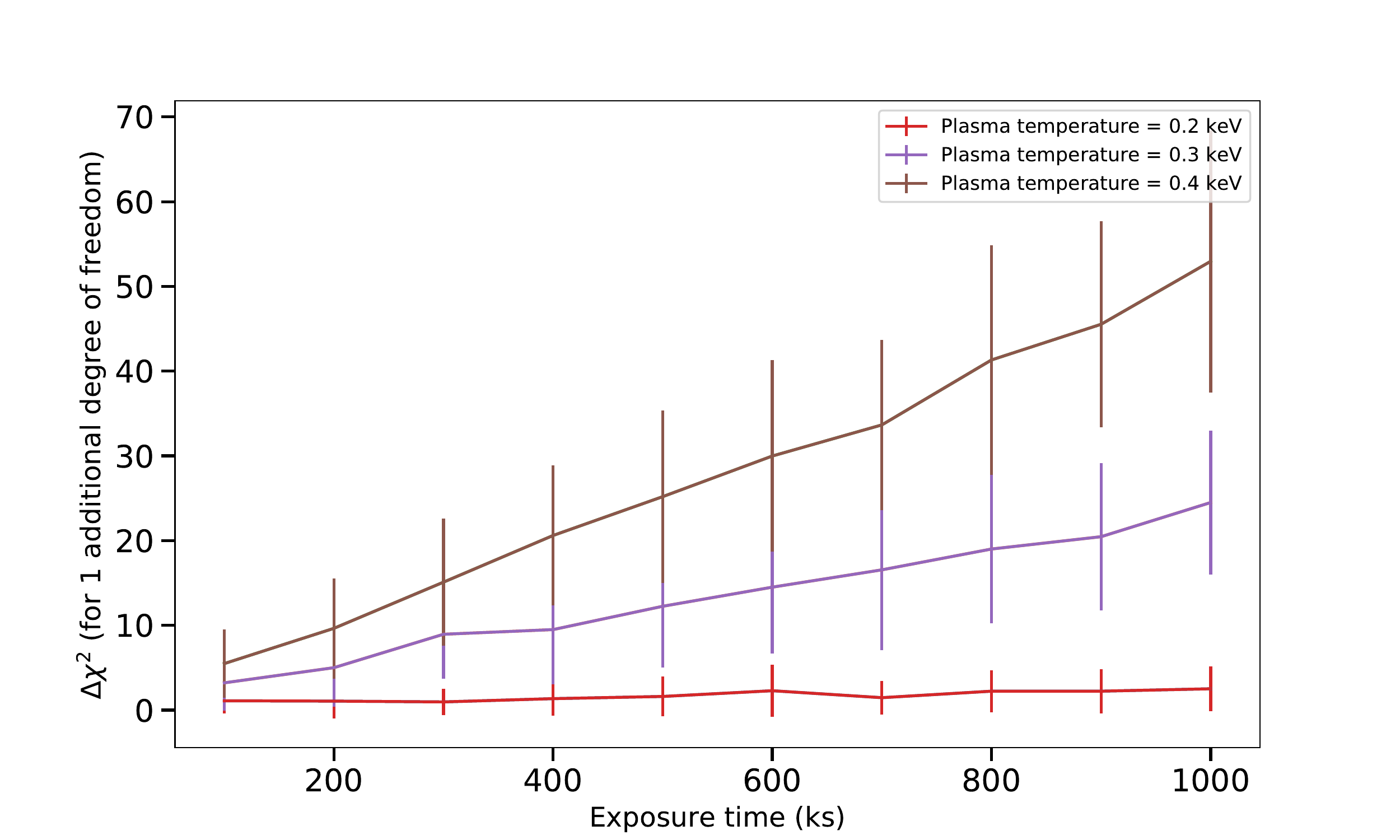}
	  \caption{The exposure time required to detect the faint ($\sim 10^{35}$ erg s$^{-1}$) X-ray emission from the bubble nebula with the \textit{Athena} X-IFU micro-calorimeter. We generate mock spectra of the X-IFU for a range of different exposure times. We model the X-ray emission from the bubble nebula with a thermal plasma emission ({\tt{apec}}) component, considering three different plasma temperatures. The ULX emission is modelled with a standard multi-colour disc plus power-law component. The vertical axis indicates the difference in the chi-squared value when the mock spectrum is fitted with and without the {\tt{apec}} component. Clearly, the nebular emission can only be detected to a high significance for plasma temperatures larger than 0.2 keV, and for exposure times greater than $\sim 400$ ks. Each point represents an average over 100 iterations to eliminate statistical fluctuations in the goodness of fit.}   
    \label{radial_profiles}
  \end{minipage}\\[1em]
\end{figure*}    

In summary, we do not find any evidence for an extended X-ray nebula surrounding the ULX Ho IX X-1 after performing a spectral and photometric analysis of the source with archival \textit{Chandra} data. We estimate an upper-limit on the thermal plasma emission from the shocked wind to be $< 10^{36}$ erg s$^{-1}$ (at the 3$\sigma$ confidence level). This implies that the simulations of \citet{nebula_sim} do not under-estimate the X-ray emission from ULXBs. Our result also highlights that the soft residuals in the \textit{XMM-Newton} spectrum of Ho IX X-1 are likely to be attributed to processes spatially localised to the source, and not to an extended X-ray nebula. As discussed in section 1, there is strong evidence that the spectral residuals are associated with absorption/emission by a fast-moving outflow in at-least four other ULXs (Pinto et al. 2016,17). But, there are very weak detections of wind absorption features in the RGS spectrum of Ho IX X-1 (Kosec et al. 2018), even though it plainly shows residuals in the broadband continuum at energies below 1 keV. According to the super-critical wind model, the non-detection of wind-absorption features implies that we are viewing this source at very small inclinations, which is consistent with its very hard X-ray spectrum. However, the residuals may still arise due to emission from a partially ionised outflow at \textit{large distances} from the compact object ($\gtrsim 10^3$ gravitational radii; \citealt{residual_features}) as it expands and becomes optically thin.   

Deep X-ray observations of bubble nebulae with future instruments could provide additional insights into the kinematics of the outflow. The X-ray emissivity of the shocked wind depends on the density of material in the wind $n$ and on the gas temperature $T$, scaling as $n^2 T^{0.5}$. The simulations predict that both of these quantities are sensitive to the unshocked outflow velocity (see Figure 7 of Siwek et al. 2017). A faster unshocked wind leads to a hotter X-ray emitting plasma, causing it to radiate with a greater emissivity. At the same time, for a given kinetic wind power, a faster wind also implies a smaller mass outflow rate (and hence electron density). This acts to decrease the X-ray emissivity overall, since the latter depends more strongly on electron density rather than temperature. Siwek et al. (2017) find that the X-ray luminosity peaks at wind velocities of $\sim 0.003$c, whilst being almost negligible (i.e. $< 10^{33}$ erg s$^{-1}$) for velocities that are ten times larger. Therefore, if limits can be placed on the X-ray emission of ULXBs with more sensitive instruments, this would enable independent constraints on the unshocked outflow velocity, which is a key input parameter of magnetohydrodynamic (MHD) simulations of accretion flows around BHs and NSs. We note that this would add to the constraints already derived from the analysis of atomic features in the RGS spectra of some ULXs (Pinto et al. 2016,17).      

We investigate the viability of such an experiment by estimating the exposure time required to detect the faint nebular emission with \textit{Chandra}, and compare this to the expectations from the X-IFU micro-calorimeter on-board \textit{Athena} (\citealt{nandra}). Mock spectra of both instruments are generated using the \texttt{fakeit} tool in \texttt{xspec}. We adapt the best-fit continuum model from section 3.1 in order to account for the broadband ULX emission, which comprises two distinct multi-colour disc and power-law components. We add an {\tt{apec}} component in order to model collisionally ionised X-ray emission from the shocked wind, and assign to it a luminosity of $10^{35}$ erg s$^{-1}$. This is typical of the estimated X-ray luminosities from the shocked wind by Siwek et al. (2017). We test the significance of the {\tt{apec}} component against the broadband ULX continuum for a range of different exposure times. Since the effective areas of both \textit{Chandra} and \textit{Athena} vary quite strongly between (0.3 - 1.0) keV, we expect the detection significance to also depend on the plasma temperature. Hence, the procedure is repeated for three different temperatures (0.2 - 0.4 keV). Unfortunately, we find that even for exposure times larger than 1 mega-second, the nebular emission is not strongly detected with \textit{Chandra} (i.e. detection significance $< 2\sigma$). A statistically significant detection may be possible with the X-IFU for gas temperatures exceeding 0.3 keV and integration times larger than $\sim 400$ ks (see Figure 6). 

In deriving the estimates above, we conservatively assume that the nebula cannot be spatially resolved by the X-IFU, although this might be unrealistic. Whilst its angular resolution is expected to be ten times lower than \textit{Chandra} (i.e. $\sim$ 5 arc-seconds), this might still be sufficient to resolve bubble nebulae with diameters $\gtrsim$ (20 - 25) arc-seconds. In that case, the required detection exposure times will be smaller than shown in Figure 6 owing to a much lower contribution from the spatially point-like ULX emission. The strength of the X-ray emission is also sensitive to the age of the bubble. During the very early stages of its evolution, the shocked wind is nearly adiabatic and radiates with an approximately ten times larger X-ray luminosity than during the snow-plough phase (i.e. after $\sim 0.2$ Myrs; see Figure 3 of Siwek et al. 2017). However, the inferred ages of a majority of ULXBs are $\sim 1$ Myr (\citealt*{ULXB_discovery}), such that their X-ray luminosities would be consistent with the value we use to estimate the detection exposure times (i.e. $\sim$ 10$^{35}$ erg s$^{-1}$). Finally, we note that numerous ULXs are found to reside in star-forming and starburst galaxies with strong diffuse X-ray emission extending several kpc from their centres. This is expected to be brighter than the intrinsic X-ray emission from ULXBs, posing further problems to their detection. Therefore, dwarf galaxies like Holmberg IX with negligible amounts of diffuse emission are more realistic targets for future X-ray observations of bubble nebulae.

\section*{Acknowledgements}

The authors wish to thank the referee for helpful comments that helped improve the quality of this paper. RS gratefully acknowledges the receipt of a studentship grant from the STFC ST/N50404X/1. TPR was funded as part of the STFC consolidated grant ST/K000861/1. RS is thankful for conversations with Chris Done during some parts of this work. This research has made use of data obtained from the \textit{Chandra} observatory.   







\bsp	
\label{lastpage}
\end{document}